\documentclass{ws-procs9x6}

\newcommand{\tr}{{\mathrm{tr}}}
\newcommand{\TR}{{\mathrm{Tr}}}

\begin{document}

\title{A variational look at QCD\footnote{\uppercase{T}his work is supported by the \uppercase{F}unda\c c\~ao para a \uppercase{C}i\^encia e a \uppercase{T}ecnologia (\uppercase{P}ortugal) under contract \uppercase{SFRH/BPD/12112/2003}.}}

\author{J.~Guilherme Milhano}

\address{CENTRA, Instituto Superior T\'ecnico (IST),\\
Av. Rovisco Pais, P-1049-001 Lisboa, Portugal\\ 
E-mail: gui@fisica.ist.utl.pt
}

\maketitle

\abstracts{I discuss the applicability of variational methods to the study of non-perturbative aspects of QCD. An illustration of the capabilities of the method pioneered by Kogan and Kovner is given through the analysis of the deconfinement phase transition in gluodynamics in 3+1 dimensions.}

\section{Introduction}

There is very little doubt that Quantum Chromodynamics (QCD) is the correct theory of the strong interactions.
Understanding the properties of the vacuum sector of QCD -- phenomena such as confinement and chiral symmetry breaking -- remains, in spite of years of attempts, one of the main problems in modern quantum field theory.

Amongst the variety of approaches used to  tackle these problems, there have been several attempts to apply  versions of the  variational Rayleigh-Ritz method \cite{rev}. The crux of my argument is that such methods, in particular Kogan and Kovner's version \cite{rev,kk}, offer a possibility to analytically obtain non-perturbative dynamical information directly from QCD. Although this information is partial and incomplete, the variational approach provides significant qualitative understanding, and, to a limited extend,  quantitative estimates.

When attempting to apply the variational principle to quantum field theory (QFT) one is faced with the discussed by  Feynman \cite{vangerooge}.
In what follows I will describe the construction of a variational trial state which, to a certain extent, overcomes these difficulties, and then illustrate the variational approach's capabilities by discussing its application to the study of the deconfining phase transition in gluodynamics in $3+1$ dimensions \cite{Kogan:2002yr,bg}.

\section{The variational Ansatz}

The first criterion to be considered when constructing a trial variational state is that the trial state ought to
be general enough to allow, through variation of its parameters, the relevant physics to be spanned.

Then there is the problem of calculability.
In QFT, our ability to evaluate path integrals is, to say the least, limited.
The requirement of calculability almost unavoidably restricts the possible form of the trial state to
a Gaussian.

Another serious problem is that of ``ultraviolet modes". The main
motivation of a variational calculation in a strongly interacting
theory is to learn about the distribution of the low momentum
modes of the field in the vacuum wave functional. However, the VEV
of the energy (and all other intensive quantities) is dominated
 by contributions of high momentum fluctuations, for the
simple reason that there are infinitely more ultraviolet modes
than modes with low momentum. 
Therefore, even if one has a clear idea of how the WF at low momenta should look, if the
ultraviolet part of the trial state is even slightly incorrect the
minimization of energy may lead to absurd results. Due to the
interaction between the high and low momentum modes, there is a
good chance that the infrared (IR) variational parameters will be
driven to values which minimize the interaction energy, and have
nothing to do with the dynamics of the low momentum modes
themselves.

Finally, in gauge theories there is the additional complication  of 
 gauge invariance. Allowable wave functionals must be invariant
under the time independent gauge transformations. If one does not
impose the Gauss' law on the states exactly, one is not solving
the right problem. The QCD Hamiltonian is only defined on the
gauge invariant states, and its action on non gauge invariant
states can be modified at will. Thus, by minimizing a particularly
chosen Hamiltonian without properly restricting the set of allowed
states, one is taking the risk of  finding a  ``vacuum'' which has
nothing to do with the physical one, but is only picked due to a
specific form of the action of the Hamiltonian outside the physical subspace. 

The trial variational state proposed by Kogan and Kovner \cite{kk} explicitly satisfies Gauss' law, has the correct UV behaviour built-in, and allows for analytical calculations -- yielding non-perturbative results -- to be carried out.

In order to illustrate the capabilities of this variational method, I now turn to the analysis of the deconfinent transition in gluodynamics.

\section{Deconfinement phase transition}

In the almost 25 years since the pioneering work of Polyakov \cite{polyakov} and Susskind \cite{susskind} much effort has been devoted to attempts to understand both the basic physics and quantitave features of the deconfining phase transition of QCD.

The high temperature phase is becoming well understood, and is widely believed to resemble a plasma of almost free quarks and gluons.
However, the transition region, $T_c<T<2 T_c$, is very poorly understood. This region is the most interesting since it is there that the transition between `hadronic' and `partonic' degrees of freedom occurs.

The study of the transition region is a complicated and inherently non-perturbative problem which has mostly evaded treatment by analytical methods.
Recently, the method introduced  by Kogan and Kovner \cite{kk} was applied to the study of the phase transition in  $SU(N)$ gluodynamics  \cite{Kogan:2002yr,bg}.
Following  \cite{Kogan:2002yr,bg}, we minimize the relevant thermodynamic potential at finite temperature, i.e. the Helmholtz free energy, on a set of suitably chosen trial density matrix functionals. 
We consider density matrices which, in the field basis, have Gaussian matrix elements and where gauge invariance is explicitly imposed by projection onto the gauge-invariant sector of the Hilbert space
\begin{multline} \label{qcdans}
\rho [A,A^{'}] =   \int DU \; \exp\bigg\{-\frac{1}{2}\int_{x,y}
\Big[ A_{i}^{a} (x) G^{-1 ab}_{ij}(x,y) A_{j}^{b}(y)\\
+ A^{'Ua}_{i} (x) G^{-1 ab}_{ij}(x,y) A^{'Ub}_{j}(y)
- 2 A_{i}^{a} (x) H^{ab}_{ij}(x,y) A^{'Ub}_{j} (y) \Big] \bigg\}\, ,
\end{multline}
where $\int_{x,y}=\int d^3 x\, d^3 y$, $DU$ is the $SU(N)$ group-invariant measure, and under an $SU(N)$ gauge transformation $U$
\begin{gather} \label{gt}
A^{a}_{i} (x) \rightarrow A^{U a}_{i} (x) = S^{ab} (x) A^{b}_{i} (x) + \lambda^{a}_{i} (x)\, ,
\end{gather}
with 
\begin{equation}
S^{ab} = \frac{1}{2} \mathrm{tr} ( \tau^a U^{\dag} \tau^b U )\, ,\qquad
\lambda^{a}_{i} =  \frac{i}{g} \mathrm{tr} ( \tau^a U^{\dag} \partial_i  U )\, ,
\end{equation} 
and 
$\frac{\tau^a}{2}$ form an $N \times N$ Hermitian representation of $SU(N)$:
$[ \frac{\tau^a}{2} , \frac{\tau^b}{2} ] = i f^{abc} \frac{\tau^c}{2}$ with normalization 
$\mathrm{tr} ( \tau^a \tau^b ) = 2 \delta^{ab}$.

We take the variational functions diagonal in both colour and Lorentz indices, and translationally invariant
\begin{equation}
    G^{-1ab}_{ij} (x,y) = \delta^{ab} \delta_{ij}
    G^{-1}(x-y)\, ,\quad
H^{ab}_{ij} (x,y) = \delta^{ab} \delta_{ij}
    H(x-y)\, .
\end{equation}
Further, we split the momenta into high and low modes with $k \lessgtr M$ and restrict
the kernels $G^{-1}$ and $H$ to the one parameter momentum space forms
\begin{equation} \label{rkers}
G^{-1}(k) = \left\{ \begin{matrix} M, \; k<M \\k, \; k>M \end{matrix}\right. \, , \qquad
H (k) = \left\{ \begin{matrix} H, \; k<M \\0, \; k>M \end{matrix}\right.\, .
\end{equation}

The logic behind this choice of ansatz is the following. At finite
temperature we expect $H(k)$ to be roughly proportional to the
Bolzmann factor $\exp\{-{\mathsf E}(k)\beta\}$. In our ansatz, the role of
one particle energy is played by the variational function
$G^{-1}(k)$ and its form is motivated by the propagator of a massive scalar field, i.e. $(k^2+M^2)^{1/2}$. We will be  interested only in temperatures close to
the phase transition, and those we anticipate to be small,
$\mathsf{T_{c}}\le M$. For those temperatures one particle modes
with momenta $k\ge M$ are not populated, and we thus put $H(k)=0$.
For $k\le M$ the Bolzmann factor is non-vanishing, but small.
Further, it depends only very weakly on the value of the momentum.
With the above restrictions on the kernels,  only two variational parameters, $M$ and $H$, remain.

Importantly, the density matrix functional in eq.~(\ref{qcdans}) describes, for $H=0$, a pure state $\rho=|\Psi [A]><\Psi [A]|$ where $\Psi [A]$ are Gaussian wave functionals. For $H\neq 0$, eq.~(\ref{qcdans}) describes a mixed state with $|H|$ proportional to the entropy of the trial density matrix.

The expectation value of a gauge invariant operator in the variational state eq.~(\ref{qcdans}) is then given by
\begin{align}
\label{eq:expvalue}
    \langle {\mathcal O}\rangle_{A,U}
    & = Z^{-1}\TR (\rho {\mathcal O}) \nonumber  \\
    & = Z^{-1} \int   {\mathcal D}U{\mathcal D}A
    \:{\mathcal O}(A,A')\cdot\nonumber\\
	&\quad\cdot \exp \bigg\{ -\frac{1}{2} \Big[ A G^{-1} A +  {A'}^U G^{-1}{A'}^U
         - 2A H {A'}^U \Big]\bigg\}\Bigg|_{A'=A}\, ,
\end{align}
where $Z$ is the normalization of the trial density matrix $\rho$, i.e.
\begin{equation}
    Z=\TR\rho 
	= \int {\mathcal D}U{\mathcal D}A \,
	\exp \bigg\{
    -\frac{1}{2} \Big[
    A G^{-1} A +  {A}^U G^{-1}{A}^U
    -2 A H {A}^U \Big]\bigg\} 
\end{equation}

To evaluate the above expressions we first perform, for fixed $U(x)$, the Gaussian integration over the vector potential $A$. For $Z$ we get, in leading order in $H$,
\begin{multline}
    Z=  \int {\mathcal D}U \exp \bigg\{
    -\frac{1}{2}\lambda\Big(
    \frac{G^{-1}}{2} + \frac{H}{4} (S+S^T)
    \Big)\lambda
+\frac{3}{4} HG\:\tr (S+S^T)\bigg\}\, .
\end{multline}

We now integrate out the high momentum, $k^2>M^2$,  modes of $U$ perturbatively to one-loop order.
This effects a renormalization group transformation on the low modes, replacing the bare coupling $g^2$ by the running coupling $g^2(M)$ \cite{kk,brko}.
To one-loop accuracy, the coupling $g^2(M)$ runs identically to the Yang-Mills coupling \cite{brko}.

The normalization $Z$ can be then interpreted as the generating functional 
\begin{equation}
    Z=\TR\rho =  \int {\mathcal D}U e^{- {\mathcal S}(U)}
\end{equation}
for an effective non-linear $\sigma$-model for the low momentum modes ($k^2<M^2$) defined by the action
\begin{multline}
    \label{eq:action}
    {\mathcal S}(U) =\frac{M}{2 g^2} \tr( \partial U\partial U^\dagger)
   - \frac{H}{8 g^2}\tr\Big[(U^\dagger \partial U - \partial U^\dagger U)
    (\partial U U^\dagger - U\partial U^\dagger) \Big]\\
    \quad- \frac{1}{4\pi^2} H M^2 \tr U^\dagger \tr U\, ,
\end{multline}
where $U$ independent pieces have been dropped.

The matrix $U$ plays the same role as Polyakov's loop $P$ at
finite temperature --- the functional integration over $U$ projects
out the physical subspace of the large Hilbert space on which the
Hamiltonian of gluodynamics is defined.

This $\sigma$-model has a  phase transition \cite{kk} at the critical point (for $\Lambda_{QCD} = 150$Mev, $N=3$ and  with the one-loop Yang-Mills $\beta$ function)
\begin{equation}
M_c=\Lambda_{QCD}e^{\frac{24}{11}}=8.86\Lambda_{QCD}=1.33 \mathrm{Gev}\, .
\label{mc}
\end{equation}
For $M<M_c$, the $\sigma$-model is in a disordered, $SU(N)_L\otimes SU(N)_R$ symmetric, phase with massive excitations and where $\langle U\rangle = 0$. Since $U$ is  the Polyakov loop, this corresponds to a confined state. When $M>M_c$, the $\sigma$-model is in an ordered, $SU(N)_V$ symmetric, phase with massless Goldstone bosons for which $\langle U\rangle \neq 0$, corresponding to a deconfined state.
With this analysis we have established a correspondence between the $\sigma$-model phase transition and the deconfinement transition in $SU(N)$ gluodynamics.

In fact, this correspondence can be argued in rather general terms.
Let us ask ourselves what would happen if we did not restrict $H$
to be small, and more generally did not restrict the functional
forms of $G(k)$ and $H(k)$ in our variational ansatz. We could
still carry on our calculation for a while. Namely we would be
able to integrate over the vector potentials in all averages, and
would reduce the calculation to a consideration of some non-linear
$\sigma$-model of the $U$-field. This $\sigma$-model quite
generally will have a symmetry breaking phase transition as the
variational functions $G(k)$ and $H(k)$ are varied. Since at this
transition the Polyakov loop $U$ changes its behavior, the
disordered phase of the $\sigma$-model corresponds to the
confining phase of the Yang-Mills theory, while the ordered phase
of the $\sigma$-model represents the deconfined  phase. Thus, in
order to study deconfinement in the $SU(N)$ Yang-Mills theory, we
should analyze the physics of each $\sigma$-model phase as accurately
as possible and calculate the transition scale $M_c$ (or rather
$G_c(k)$). We then calculate the free energy of the $\sigma$-model
in each phase at temperature $T$ and extract the minimal free
energy. The deconfinement transition occurs at the temperature for
which the free energies calculated in the ordered and disordered
phases of the sigma model coincide.

The Helmholtz free energy ${\sf F}$ of the density matrix
$\rho$ is given by
\begin{equation}
    {\mathsf F} = \langle {\mathsf H} \rangle
     - {\mathsf T}   \langle{\mathsf S} \rangle\, ,
    \end{equation}
where ${\sf H}$ is the standard Yang-Mills Hamiltonian
\begin{equation}
{\mathsf H}= \int d^{3}x \left[{1\over 2}E^{a2}_i+{1\over
2}B^{a2}_i\right]\, , \label{ham}
\end{equation}
with
\begin{align}
E^a_i(x)&=i{\delta\over \delta A^a_i(x)}\, , \nonumber \\
B^a_i(x)&={1\over 2}\epsilon_{ijk}
\{\partial_jA_k^a(x)-\partial_kA^a_j(x)
+gf^{abc}A_j^b(x)A_k^c(x)\}\, ,
\end{align}
${\sf S}$ is the entropy, and ${\sf T}$ is the
temperature.

Thus
\begin{equation}
\label{eq:freeenergy}
     {\mathsf F}
    = \frac{1}{2}\Big(
    \TR ({E}^2\varrho)
    +\TR ({B}^2\varrho)
    \Big)
    + {\mathsf T}\cdot\TR( \varrho \ln\varrho) \, .
\end{equation}

In the disordered phase, no progress seems possible without restricting the arbitrary kernels.  Following \cite{Kogan:2002yr},
we adopt the forms eq.~(\ref{rkers}).
For small $H$, we consider only the first non-trivial order in $H$, that is a term of $o(H\ln H)$ in the entropy. This term can be written as a product of  left $SU(N)$ and  right $SU(N)$ currents and does, therefore, vanish in the disordered, $SU(N)_L\otimes SU(N)_R$ symmetric, phase \cite{Kogan:2002yr}. The remaining contribution to the free energy, the average of the Hamiltonian, is evaluated in the mean field approximation \cite{kk}. The free energy is minimized for $M=M_c\simeq 1.33 \mathrm{GeV}$
\begin{gather} \label{minlf}
F_{dis} = - \frac{N^2 M_{c}^{4}}{30 \pi^2}\, .
\end{gather}

The simplest option to evaluate the free energy in the disordered phase is to use perturbation theory. Perturbation theory is certainly appropriate for large enough values of $M$, where the expectation value of the $U$ field is close to unity. From numerical studies \cite{kogut} it is known that the transition occurs when the expectation value of $U$ is greater than $0.5$. We can thus expect perturbation theory to be qualitatively reliable down to the transition point. 
In the leading order perturbation theory approximation to the ordered phase of the  $\sigma$-model, however, minimisation with respect to arbitrary kernels $G^{-1}$ and $H$ for both high and low modes is possible. Further, the analysis can be carried out to all orders in the thermal disorder kernel $H$.

In this approximation, the $U$ matrices can be parameterized in the standard exponential form and expanded in the coupling $g$
\begin{gather}
U = \exp\bigg\{ig \varphi^a \frac{\tau^a}{2}\bigg\} = 1 + ig \varphi^a \frac{\tau^a}{2} + \dots  
\end{gather}
Hence at leading order one can take
\begin{equation}
U \simeq 1\, , \qquad 
\partial_i U \simeq  ig \partial_i \varphi^a \frac{\tau^a}{2}. 
\end{equation}
Thus, the gauge transformations (\ref{gt}) reduce to
\begin{gather}
A^{a}_{i} \rightarrow A^{a}_{i} - \partial_i \varphi^a  
\end{gather}
and the Hamiltonian (\ref{ham}) reduces to 
\begin{gather}
\mathcal{H} = \frac{1}{2} \left[ E^{a2}_{i} + (\epsilon_{ijk} \partial_j A^{a}_{k})^2 \right]\, . 
\end{gather}
These last two equations describe the theory $U(1)^{N^2-1}$: in the leading order of $\sigma$-model perturbation theory, the 
$SU(N)$ Yang--Mills theory reduces to the $U(1)^{N^2-1}$ free theory.
Moreover, the density matrix eq.~(\ref{qcdans}) becomes Gaussian again, because the gauge transformations are linear:
\begin{multline} \label{gans}
\rho [A,A^{'}] = \int D\varphi \; 
\exp\bigg\{-\frac{1}{2} \Big[ A G^{-1} A
+ (A' - \partial \varphi) G^{-1} (A' - \partial \varphi) \\
- 2 A H (A' - \partial \varphi) \Big] \bigg\}\, . 
\end{multline}

The theory of $N^2-1$ $U(1)$ free fields in $3+1$ dimensions 
is completely tractable; the variational analysis for the $U(1)$ theory
(with Gaussian ansatz (\ref{gans})) was discussed in \cite{Gripaios:2002xb}. The free energy in momentum space in terms of the arbitrary kernels
$G^{-1}$ and $H$ is
\begin{multline}
{\mathsf F} = \frac{N^2-1}{2} \int \frac{d^3p}{(2\pi)^3}
\Bigg[  G^{-1}(1 + GH) + p^2 G (1 - GH)^{-1} \\
- 4{\mathsf T}  \left(  \ln \bigg[ \frac{GH}{\xi}\bigg]
 - \ln \bigg[ \frac{\eta}{GH}\bigg]
 \cdot \frac{\eta}{\xi}  \Bigg) \right]\, ,
\end{multline}
where $\eta=1- (1-(GH)^2)^{1/2}$ and $\xi=(1-(GH)^2)^{1/2}-(1-GH)$.
The kernels which minimize the free energy are
\begin{equation} \label{mkers}
G^{-1} = p 
\left( \frac{1+ e^{-\frac{2p}{T}}}{1 - e^{-\frac{2p}{T}}}\right)\, , \qquad
H = 2p
 \left( \frac{e^{-\frac{p}{T}}}{1 - e^{-\frac{2p}{T}}}\right)
\end{equation}
and the minimal value of the free energy at temperature $T$ is
\begin{gather}
F = - \frac{\pi^2 N^2 T^4}{45}.
\end{gather}

So we see that the free energy of $SU(N)$ is minimised with $M=M_c$ in the disordered phase of the $\sigma$-model for temperatures from zero up to
a temperature $T_c$ where
\begin{gather} 
F = - \frac{N^2 M_{c}^{4}}{30 \pi^2} =  - \frac{\pi^2 N^2 T_{c}^{4}}{45}\, ,
\end{gather}
which in turn implies 
\begin{gather}
T_c = \left( \frac{3}{2}\right)^{1/4} \frac{M_c}{\pi} \simeq 470 \mathrm{MeV}\, .
\end{gather}

Although the actual value of the transition temperature is
considerably larger than the lattice estimate it makes more sense to look at dimensionless quantities.
In particular, if we identify $2M_c$ with the mass of the lightest
glueball  \cite{Gripaios:2002bu} (see also the discussion in \cite{rev}), we find
\begin{gather} \label{ratio}
\frac{{\mathsf T}_c}{2M_c} = \frac{1}{2\pi}\left(\frac{3}{2}\right)^{1/4} \simeq 0.18\,.
\end{gather}
This is in excellent agreement with the lattice estimate for
$SU(3)$ pure gauge theory \cite{Teper:1998kw}.

\section{Conclusions}

In 3+1 dimensional gluodynamics, this variational method gives
results which on the qualitative level at least, conform with our
intuition about the structure of the ground state.
We find a first order phase transition which
corresponds to the Polyakov loop acquiring a non-zero average.
Although we have not calculated the string tension directly, the
behavior of the Polyakov loop is very much indicative that this is
indeed the deconfining phase transition. The value of the critical
temperature (in units of glueball mass) we find is in good
agreement with lattice results. We also found that in the low
temperature phase the entropy remains zero all the way up to the
transition temperature. This is a rather striking result, which
has not been built into our variational ansatz, but rather emerged
as the result of the dynamical calculation.

An important lesson we learned 
is that the projection of the Gaussian trial state onto the gauge
invariant Hilbert subspace dictates most, if not all, of the
important aspects of the non-perturbative physics. It was
absolutely essential to perform the projection non-perturbatively,
fully taking into account the contribution of the overlap between
gauge rotated Gaussians into the variational energy prior to
minimization.
We have seen that from the point of view of the effective
$\sigma$-model the energy is minimized in the disordered phase. In
other words, the low momentum fluctuations of the field $U$ are
large, unlike in the perturbative regime, where $U$ is close to a
unit matrix. From the point of view of the trial wave functional,
this means that the off-diagonal contributions, coming from the
Gaussian WF gauge rotated by a slowly varying gauge
transformation, are large. It is these ``off diagonal" contributions
to the energy that lowered the energy of the best trial state
below the perturbative value. In the low temperature phase the
vanishing of the entropy was also a direct consequence of the
effective $\sigma$-model being in the disordered phase, and thus
of the non-perturbative nature of the gauge projection.

This variational method appears to be a good candidate for a useful calculational scheme for strongly interacting gauge
theories. 
However, many outstanding questions remain. Is the best variational state
confining? How do we calculate the interaction potential between
external sources? How do we understand better the relation between
the variational parameter and the glueball masses? Can we extend
the Ansatz to include (massless) fermions?

Personally, I  believe that these results are genuine and that there is
enough scope for further development of the approach which
warrants continuing active investigations.

\section*{Acknowledgments}
This talk is based upon work done in collaboration with Ben Gripaios , Alex Kovner and Ian Kogan.


\begin{thebibliography}{0}

\bibitem{rev} A. Kovner and J. G. Milhano, in M. Shifman, J. Wheater and A. Vainstein (eds.): {\em Circumnaviagting Theoretical Physics} (World Scientific, Singapore, 2004), [hep-ph/0406165].

\bibitem{kk} I. Kogan and A. Kovner, Phys. Rev.
D52, 3719 (1995); e-Print Archive: hep-th/9408081;

\bibitem{vangerooge} R. Feynman, in Wangerooge 1987, Proceedings,
{\em Variational calculations in quantum field theory}, L. Polley and
D. Pottinger, eds (World Scientific, Singapore, 1988).

\bibitem{Kogan:2002yr}
I.~I.~Kogan, A.~Kovner and J.~G.~Milhano,
JHEP {\bf 0212}, 017 (2002)
[arXiv:hep-ph/0208053].

\bibitem{bg}
B.~M.~Gripaios and J.~G.~Milhano,
Phys. \ Lett. \ B {\bf 564} (2003) 104
[arXiv:hep-ph/0302172].




\bibitem{polyakov} A.~Polyakov, Phys. \ Lett. \ B {\bf 72} (1978) 477

\bibitem{susskind} L.~Susskind, Phys. \ Rev. \ {\bf 20} (1979) 2610





\bibitem{Teper:1998kw}
M.~J.~Teper,
arXiv:hep-th/9812187.

\bibitem{brko} W. E. Brown and I. I. Kogan,
Int.J.Mod.Phys.A14:799,1999; e-Print Archive: hep-th/9705136;

\bibitem{kogut}  J. B. Kogut, M. Snow and M. Stone  Nucl.Phys.B200:211,1982

\bibitem{Gripaios:2002bu}
B.~M.~Gripaios,
Int.\ J.\ Mod.\ Phys.\ A {\bf 18} (2003) 85
[arXiv:hep-ph/0204310].

\bibitem{Gripaios:2002xb}
B.~M.~Gripaios,
Phys.\ Rev.\ D {\bf 67}, 025023 (2003)
[arXiv:hep-th/0211104].




\end{thebibliography}
\end{document}